\documentclass[aps,prc,twocolumn]{revtex4}
\usepackage{graphicx}
\usepackage{amsmath,bbm}
\usepackage{amssymb,bm}
\usepackage{array}
\begin{document}
\title{Electrical Conductivity of an Anisotropic Quark Gluon Plasma : A Quasiparticle Approach}
\author{P. K. Srivastava\footnote{$prasu111@gmail.com$}}
\author{Lata Thakur\footnote{$lata1dph@iitr.ac.in$}}
\author{Binoy Krishna Patra\footnote{$binoyfph@iitr.ac.in$}}
\affiliation{$^1$ Department of Physics, Indian Institute of Technology 
Roorkee, Roorkee~247667, INDIA}
\begin{abstract}
The study of transport coefficients of strongly interacting 
matter got impetus after the discovery of perfect fluid 
ever created at ultrarelativistic heavy ion collision experiments. 
In this article, we have calculated one such coefficient viz. electrical 
conductivity of the quark gluon plasma (QGP) phase which exhibits a 
momentum anisotropy.  Relativistic Boltzmann's kinetic equation has been solved in the 
relaxation-time approximation to obtain the electrical conductivity. 
We have used the quasiparticle description to define the basic 
properties of QGP. We have compared our model results with 
the corresponding results obtained in different lattice as well as 
other model calculations. Furthermore, we extend our model to calculate 
the electrical conductivity at finite chemical potential.
\\

 PACS numbers: 12.38.Mh, 12.38.Gc, 25.75.Nq, 24.10.Pa
\end{abstract}

\maketitle 
\section{Introduction}
\noindent
Transport coefficients are of particular interest to quantify 
the properties of strongly interacting matter created at 
relativistic heavy ion collisions (HIC) and these coefficients can be 
instrumental to study the 
critical properties of QCD medium. The fluctuations or external 
fields cause the system to depart from its equilibrium and a 
non-equilibrium system has been created for a brief time. 
The response of the system to such type of fluctuations or 
external fields is essentially described by transport 
coefficients {\em eg}. the shear and bulk viscosities, the speed of 
sound etc. In recent years a somewhat surprising result in 
the quark gluon plasma (QGP) story has occurred when the 
practitioners in this field tried to satisfy the collective 
flow data as obtained in collider experiments. In order to get 
the required collective flow in the framework of viscous 
hydrodynamics, the value of shear viscosity to entropy density 
ratio ($\eta/s$) comes out to be very small~\cite{roma,heinz,kotun}. 
The tiny value of $\eta/s$ indicates the discovery of most perfect 
fluid ever created in laboratory. This perfect fluid is described 
as strongly interacting quark gluon 
plasma~\cite{lee,guy,shuryak,hirano}. 
Thus the study of various transport coefficients is a powerful 
tool to really understand the behaviour of the matter produced 
in the ultra relativistic heavy ion collision (uRHIC) experiments at 
RHIC and LHC.\\

Recently electrical conductivity has gained a lot of interest 
due to the strong electric field created in the collision zone 
of uRHIC experiments~\cite{vgreco,fraile,greif}. 
It has been observed that strong electric and magnetic fields are 
created in peripheral heavy ion collisions whose strength are 
roughly estimated as $eE=eB=m_{\pi}^{2}$(where $m_{\pi}$ is 
the mass of the pion) within proper time $1-2$ fm/c~\cite{tuchin}. 
This large electrical field can significantly affect the 
behaviour of the medium created in these collisions and the
effect depends on the magnitude of electrical conductivity ($\sigma_{\rm{el}}$) 
of the medium. $\sigma_{\rm{el}}$ is responsible for the production of electric 
current generated by the quarks in the 
early stage of the collision. The value of $\sigma_{\rm{el}}$ would 
be of fundamental importance for the strength of Chiral Magnetic 
Effect~\cite{fuku} which is a signature of CP-violation in 
strong interaction. Further the electrical field in mass asymmetric 
collisions (e.g. $Cu$-$Au$ collisions etc.) has overall a preferred 
direction and thus generating a charge asymmetric flow whose 
strength is directly related to $\sigma_{\rm{el}}$~\cite{hongo}. 
Furthermore, $\sigma_{\rm{el}}$ is related with the emission rate 
of soft photons~\cite{kapusta} accounting for their raising 
spectra~\cite{turbide,linnyk}. Despite of the importance of 
electrical conductivity, it has been studied rarely in the 
literature for the QGP phase.\\

With the discovery of ``most perfect fluid ever generated",
another important observation has been made that this fluid 
possesses momentum-space anisotropies in the local rest frame 
(LRF)~\cite{strick,strick1}. This has important implications 
for both dynamics and signatures of the QGP. Earlier it has 
been assumed {\em a priori} in the ideal hydrodynamics that the 
QGP is completely isotropic. However, recently dissipative 
hydrodynamics helps us to understand that the QGP created in 
ultrarelativistic heavy ion collisions has different 
longitudinal and transverse pressure. It has been shown 
heuristically in first-order Navier Stokes viscous 
hydrodynamics that the ratio of longitudinal pressure over 
the transverse pressure is: $P_{L}/P_{T}=(3\tau T-16\bar{\eta})/(3\tau T+8\bar{\eta})$, 
where $\bar{\eta}=\eta/s$~\cite{strick}. Using the RHIC-like initial 
condition the value of $P_{L}/P_{T}$ comes out equal to $0.5$ 
and for LHC-like initial condition this ratio takes the value 
as $0.35$~\cite{strick}. It has also been shown that there 
exists an anisotropy in $P_{L}$ versus $P_{T}$ in second-order Israel-Stewart viscous hydrodynamics. 
Several other groups who study the early-time dynamics of QCD within 
AdS/CFT framework have also shown the early-time pressure anisotropies and 
quote $P_{L}/P_{T}=0.31$ or smaller~\cite{heller, vander}. 
In colour-glass condensate framework, the practitioners have found that the 
timescale for isotropization in classical Yang-Mills simulations 
is very long~\cite{mac, iancu}. From these observations and 
findings one can certainly assumes that the momentum anisotropy 
produced in the medium created in heavy ion collisions lasts 
for at least $\tau \leq 2~fm/c$. Thus it is crucial to incorporate these momentum-space 
anisotropies in any phenomenological studies specifically 
for transport coefficients.\\

In this article our main motivation is to calculate the 
electrical conductivity of an anisotropic QGP phase using 
the Relativistic Boltzmann's kinetic equation. We have used 
the quasiparticle description to define the basic properties 
of QGP since earlier we have shown that 
quasiparticle description provides the proper and realistic 
thermodynamical and transport behaviour of QGP phase~\cite{pks,pks1}. 
We want to provide the correct temperature dependence of
$\sigma_{\rm{el}}$ since it is not yet established. Lattice calculations have 
obtained $\sigma_{\rm{el}}$ at few values of temperature and 
these estimates vary widely~\cite{aamato,ding,aarts,gupta}. Thus it is 
necessary to provide a proper temperature dependence to electrical 
conductivity. We will also extend our calculation at 
finite quark chemical potential ($\mu_q$). This is 
important since there is no guidance at finite $\mu_q$ due to 
severe limitation of lattice calculations in this region. 
Very recently Berrehrah and collaborators have shown the 
variation of $\sigma_{\rm{el}}/T$ with respect to temperature 
at small but finite chemical potential in dynamical 
quasiparticle model (DQPM)~\cite{berre}.\\

 Rest of the article is organized as follows : firstly we 
 present the calculation of electrical conductivity in the 
 isotropic case. In second subsection we introduce 
 a momentum-space anisotropy in the distribution function of 
 quarks and antiquarks and then derive the expression of 
$\sigma_{\rm{el}} $ for this anisotropic hot and/or dense QCD 
medium using Relativistic Boltzmann's approach. 
 In the third subsection we will provide a brief introduction 
 about quasiparticle model. We further explain the importance 
 of quasiparticle description of QGP in comparison to ideal description
. Later we demonstrate results obtained in our model  
 at zero $\mu_q$ and their comparison with the corresponding 
 results obtained in lattice as well as in phenomenological calculations. 
 We have also extended our model calculation at finite 
 $\mu_q$. Finally we will give the summary and the 
 conclusions drawn from this work.

\section{Description of Model}
\subsection{Electrical Conductivity for Isotropic System}
The electric conductivity represents the 
response of the system to applied electric field. According 
to Ohm's law, the spatial electric 
current ($ {\bf J} $) is directly proportional to the 
longitudinal component of the electric field $ {\bf E} $ :
\begin{equation}
{\bf J}= \sigma_{\rm{el}} {\bf E},
\end{equation}
where the proportionality coefficient is known as electrical 
 conductivity ($\sigma_{\rm{el}}$). One can write down the expression for four 
 current ($ J^{\mu} $) in a covariant form as:
\begin{eqnarray}
J^{\mu}=  \int \frac{d^{3}p}{(2 \pi)^3E} p^{\mu}\lbrace q 
~g~f(x,p)- {\bar q} ~g~{\bar f} (x,p) \rbrace,
 \label{current}
\end{eqnarray}
where $ f(x,p)( {\bar f(x,p)} $) and q (${\bar q} $) are the distribution 
function and electric charge for quark (anti-quark), respectively. Further $g$ is the degeneracy factor. Let us first 
assume the case of vanishing chemical potential i.e., 
$ \mu_q =0 $. In this case quark and anti-quark distribution 
functions will become identical, so Eq. (\ref{current}) 
takes the following form:
\begin{equation}
J^{\mu}= 2q_{{}_f} g_{{}_f} \int \frac{d^{3}p}{(2 \pi)^3E}p^{\mu}f(x,p).
\end{equation}
In the presence of infinitesimal external disturbance, the change in four current is : 
\begin{equation}
\Delta J^{\mu}= 2q_{{}_f} g_{{}_f} \int \frac{d^{3}p}{(2 \pi)^3E}p^{\mu} \delta f(x,p),
\label{delj}
\end{equation}
where $\delta f$ is the change in distribution function due to external disturbance.
One can obtain the $ \delta f(x,p) $ by using the 
Relativistic Boltzmann Transport (RBT) equation,
which is given by~\cite{Yagi, Cercignani}:
\begin{equation} 
\label{Boltzmann_eq}
p^{\mu}\partial_{\mu} f(x,p) + q F^{\alpha\beta}p_{\beta} 
\frac{\partial}{\partial p^{\alpha}} f(x,p) = {\cal C}[f],
\end{equation}
where $F^{\alpha\beta}$ is the electromagnetic field strength 
tensor and  $C[f]$ is 
the collision integral, which in the relaxation-time 
approximation is given by :
\begin{equation}
{\cal C}[f] \simeq -\frac{p^{\mu}u_{\mu}}{\tau} (f-f^{0}), 
\, \, (f-f^{0})=\delta f
\label{cf}
\end{equation}
where $ \tau $ is the relaxation time and $ f^{0} $ is the equilibrium  
distribution function.  In this approximation, RBT equation becomes:
\begin{equation} 
p^{\mu}\partial_{\mu} f(x,p) + q F^{\alpha\beta}p_{\beta} 
\frac{\partial}{\partial p^{\alpha}} f(x,p) = -\frac{p^{\mu}u_{\mu}}{\tau} \delta f
\label{RBT}
\end{equation}
where $ u^{\mu}$ is the fluid four velocity and in the 
local rest frame i.e., $ u^{\mu}=(1, {\bf 0}) $. The equilibrium 
quark distribution function at $\mu_q=0$ has the following form :
\begin{equation}
f^{0}(x,{\bf p};T)=\frac{1}{e^{E/T}+1}, 
\end{equation}
where $ E=\sqrt{{\bf p}^{2}+m^{2}}$. \\

Since we are interested only in electric field 
components of $F^{\alpha\beta}$, we take only two terms: 
$ F^{0i}=-{\bf E}~\, \, and \, \, F^{i0}={\bf E} $. Thus 
the RBT equation (\ref{RBT}) becomes:
\begin{equation}
-q\left( p_0 {\bf E} \cdot \frac{\partial f^{0}}{\partial {\bf p}} - {\bf E}\cdot {\bf p} \frac{\partial f^{0}}{\partial p^0} \right) = -\frac{p^0}{\tau} \delta f,
\label{RBT1}
\end{equation}
where $ \frac{\partial f^{0}}{\partial {\bf p}} $ can be solved by 
using the chain-rule and after differentiation we get the value of $\delta f  $ as:
\begin{equation}
\delta f= q\tau \frac{{\bf E} \cdot {\bf p}}{p^{0}T} f^{0}(1-f^{0})
\end{equation}
By substituting $ \delta f $ in Eq. (\ref{delj}) we obtain the 
expression for $ \sigma_{\rm{el}} $:
\begin{equation}
\sigma_{\rm{el}}^{\rm{iso}} = \frac{2}{T} \sum_f g_{{}_f}q_{{}_f}^{2}\int\frac{d^3p}{(2\pi)^3} 
\frac{{p}^2}{E^2}  \tau_{{}_f} \times f_{{}_f}^{0}(1-f_{{}_f}^{0}),
\label{sigiso}
\end{equation}
where the subscript $f$ implies summation over the flavors. Here we have 
taken up, down and strange flavors only.\\

It is important to provide a connection between the calculation 
of electrical conductivity by thermal field theory and in Boltzmann's 
kinetic approach. In an equilibrated system having volume $V$ and 
temperature $\beta^{-1}$, the zero frequency Green-Kubo~\cite{green,kubo} 
formula for the electrical conductivity is given by the current-current
autocorrelation
\begin{equation}
\sigma_{\rm{el}}=\beta V\int_{0}^{\infty}\langle j_{i}(0)j_{i}(t)\rangle dt,
\end{equation}
where the repeated spatial-indices, $i$ do not imply
summation.  In the local rest frame of fluid, the electric current density 
can be read as :
\begin{equation}
j^{l}(t)=\frac{1}{V~N_{test}}\sum_{k=1}^{M}q_{k} 
\sum_{i=1}^{N_{k}}\left(\frac{p_{i}^{l}}{p_{i}^{0}}\right)_{t},
\end{equation}
where $M$ is the number of particle-species and $N_k$ is the number of
particles of $k$-th species.
For all equivalent directions ($l$), $j_l(t)$ was obtained from 
Boltzmann Approach to  Multi-parton Scatterings
and then the autocorrelation function, $C(t)$ of the
electric current density in equilibrium has been 
extracted~\cite{greif}. For example, the variance, $C(0)$ has been
computed analytically 
\begin{equation}
C(0) =\frac{1}{3V}\sum_{k=1}^{M}q_{k}^{2}n_{k},
\end{equation}
which finally gives the expression for $\sigma_{\rm{el}}$ :
\begin{equation}
\sigma_{\rm{el}}=\frac{1}{3T}\sum_{k=1}^{M}q_{k}^{2}n_{k}\tau ~,
\end{equation}
which is nothing but the non-relativistic Drude's formula to calculate the
$\sigma_{\rm{el}}$. Similarly one can show that the expression for 
$\sigma_{\rm{el}}$ obtained by solving 
relativistic Boltzmann kinetic equation (see Eq. (11)) can also be 
approximated as Drude's formula assuming small electric field and no cross 
effects between heat and electrical conductivity~\cite{greif}. \\

At finite quark-chemical potential ($ \mu_q \neq 0 $), the 
distribution function for quark and anti-quark will be different. For quarks :
\begin{equation}
f^{0}_f(x,{\bf p};T,\mu_q)=\frac{1}{e^{(E-\mu^q_f)/T}+1},
\end{equation}
and for antiquarks : 
\begin{equation} 
{\bar f^{0}_f}(x,{\bf p};T,\mu_q)=\frac{1}{e^{(E+\mu^q_f)/T}+1}.
\end{equation}
One can then generalize the expression 
for electrical conductivity in an isotropic medium for 
$ \mu_q \neq 0 $ as follows :
\begin{eqnarray}
\sigma_{\rm{el}}^{\rm{iso}}(\mu_q \neq 0) &=& \frac{1}{T} \sum_f g_{f} q_{f}^{2}
\int \frac{d^3p}{{(2\pi)}^3} \frac{{p}^2}{E^2} \nonumber\\
&\times&\left[\tau_{{}_f} f_{{}_f}^{0}(1-f_{{}_f} ^{0})+ \tau_{\bar {{}_ f}}
{\bar {f_{{}_f}}^{0}}(1-{\bar f_{{}_f}^{0}})\right].
\label{sigiiso}
\end{eqnarray}
\subsection{Electrical Conductivity for Anisotropic 
System $ \xi \neq 0 $} The introduction of an anisotropic 
distribution function is needed to properly describe the QGP 
created in heavy ion collisions~\cite{dum1,dum2}. Partons 
are produced from the incoming colliding nuclei just after 
the collision at proper time $\tau=\tau_0\approx Q_{s}^{-1}$, 
it can be assumed that the newly produced partons follow an 
isotropic (but not necessary an equilibrium) momentum distribution. 
Here $Q_{s}$ is the gluon saturation scale. The early-time 
physics is mainly governed by the hard gluons with the 
momentum at the saturation scale which have very large occupation numbers 
of order $1/\alpha_{s}$ ($\alpha_{s}<<1$)~\cite{gribov,mueller,blaiz}.
For $\tau > Q_{s}^{-1}$, the hard gluons would  
follow the straight-line trajectories and isolate themselves 
in beam direction as if no any interaction exists. 
As a result, the longitudinal expansion causes 
the medium to become much colder in the longitudinal 
direction than the transverse direction, {\em ie.}
$p_{\perp}>> p_{z}\sim 1/\tau$ and a local momentum anisotropy 
appears.\\

The anisotropic distribution can be obtained 
by stretching or squeezing an isotropic one 
along a certain direction, thereby preserving a 
cylindrical symmetry in momentum space. In particular,
the anisotropic distribution relevant for HICs can be approximated
by removing particles with the large momentum component along 
the direction of anisotropy, ${\bf n}$ as~\cite{roma1,roma2} :
\begin{equation}
f_{\rm{aniso}} ({\mathbf p})=f_{iso}\left(\sqrt{p^{2}+\xi({\mathbf p}
\cdot {\mathbf n})^{2}}\right),
\end{equation}
where $f_{iso}$ is an arbitrary isotropic distribution 
function and $\xi$ is the anisotropic parameter and is  
generically defined as:
\begin{equation}
\xi=\frac{\langle {\bf p}_{T}^{2}\rangle}{2\langle p_{L}^{2}\rangle}-1,
\end{equation}
where $p_{L}={\mathbf p}.{\mathbf n}$ and ${\mathbf p}_{T}
={\mathbf p}-{\mathbf n}({\mathbf p}.{\mathbf n})$ are the 
components of momentum parallel and perpendicular to  
$ {\mathbf n} $, respectively. There have been significant
advances in the dynamical models used to simulate plasma evolution 
with the momentum-space anisotropies~\cite{Martinez:2010sd-12tu,
Martinez:PRC852012,Ryblewski:2010bs,Ryblewski:2012rr,Florkowski:2010cf}.
Recently two of us have investigated the effects of anisotropy on 
the quarkonia states by the leading-anisotropic
correction to the potential at T=0~\cite{lata:PRD2013,lata:PRD2014}.

If $f_{iso}$ is a thermal ideal-gas distribution and $\xi$ is 
small then $\xi$ is also related to the shear viscosity of the 
medium via one-dimensional Bjorken expansion in 
the Navier-Stokes limit~\cite{asa}::
\begin{equation}
\xi=\frac{10}{T\tau}\frac{\eta}{s},
\end{equation}
In an expanding system, non-vanishing viscosity implies 
finite-momentum relaxation rate and therefore an anisotropy 
of the particle momenta appears. For $\eta/s=0.1-0.3$ and 
$\tau T = 1-3$, one finds that $\xi=1$. \\

As we have explained, hot QCD medium due to 
expansion and non zero viscosity, exhibits a local 
anisotropy in momentum space, and the quark distribution 
function (or Fermi-Dirac distribution function) takes the following 
form for $\mu_q=0$:
\begin{equation}
f_{\rm{aniso}} (x,{\bf p};T)=\frac{1}{e^{(\sqrt{{\bf p}^{2}
 + \xi({\bf p}.{\bf n})^{2}+ m^{2}})/T}+1}.
 \label{anisof}
 \end{equation}
For weakly anisotropic systems ($\xi<<1$), one can expand the 
quark distribution function as follows :
\begin{eqnarray}
f_{\rm{aniso}} (x,{\bf p};T)&=&\frac{1}{e^{E/T}+1}-\frac{\xi}{2ET}({\bf {p\cdot n}})^{2}\frac{e^{E/T}
}{(e^{E/T}+1)^{2}},\nonumber\\
&=&f^{0}-\frac{\xi}{2ET}{({\bf p}\cdot{\bf n})}^{2}{f^0}^{2} e^{E/T},
\label{anisof1}
\end{eqnarray}
where ${\bf p}$ $\equiv$ $ (p\sin\theta\cos\phi, 
p\sin\theta\sin\phi, p\cos\theta) $ and ${\bf n}$ $\equiv$ 
$(\sin\alpha, 0, \cos\alpha)$. $ \alpha $ is the angle 
between ${\bf p}$ and ${\bf n}$. 
After substituting the anisotropic distribution function
in Eq. (\ref{sigiso}), the expression of electrical 
conductivity in anisotropic medium system is modified as:
\begin{equation}
\sigma_{\rm{el}}^{\rm{aniso}}=g_{f} \frac{2~q_{f}^{2}}{T}\int\frac{d^3p}{(2\pi)^3} 
\frac{{p}^2}{E^2}  \tau_f f_{aniso}(1-f_{aniso})
\end{equation}
On neglecting the higher-order distribution function 
(e.g., $f_{0}^{3}$ etc.) we get the expression for electrical conductivity as:
\begin{eqnarray}
\sigma_{\rm{el}}^{\rm{aniso}}&=&g_{f} \frac{2~q_{f}^{2}}{T}\int\frac{d^3p}
{(2\pi)^3} \frac{{p}^2}{E^2}  \tau_f f^{0}(1-f^{0})\\ 
&-&\xi g_{f}\frac{q_{f}^{2}}{T^2}\int\frac{d^3p}{(2\pi)^3} 
\frac{{p}^2}{E^3}  \tau_f(f^{0})^2 e^{E/T}({\bf {p\cdot n}})^{2}\nonumber
\label{siganiso1}
\end{eqnarray}
Now using the definition of $\mathbf p$ and $\mathbf n$, and integrating 
over $ \theta $ and $ \phi $, we get the electrical conductivity in anisotropic
medium, after summing over the flavours ($f$)
\begin{eqnarray}
\sigma_{\rm{el}}^{\rm{aniso}} (\mu_q=0) &=& \frac{1}{\pi^2 T} \sum_f 
g_{{}_f} q_{{}_f}^{2} \int dp \frac{{p}^4}{E^2}  
\tau_{{}_f} f_{{}_f}^{0}(1-f_{{}_f}^{0})\nonumber\\
&-&\xi \frac{1}{6 \pi^2 T} \sum_f g_{{}_f} q_{{}_f}^{2} 
\int dp\frac{{p}^6}{E^3}  
\tau_{{}_f}(f_{{}_f}^{0})^2 e^{E/T}\nonumber\\
&=& \sigma_{\rm{el}}^{\rm{iso}}- \xi A,
\end{eqnarray}
where the anisotropic term, $A$ is given by
\begin{eqnarray}
A=\frac{1}{6\pi^{2} T} \sum_f g_{{}_f} q_{{}_f}^{2} 
\int dp\frac{{p}^6}{E^3}  
\tau_{{}_f}(f_{{}_f}^{0})^2 e^{E/T}
\end{eqnarray}

Like in the isotropic medium, we can generalize the electrical 
conductivity in an anisotropic medium  for $ \mu_q\neq 0 $:
\begin{eqnarray}
\sigma_{\rm{el}}^{\rm{aniso}} (\mu_q\neq 0)&=&\frac{1}{2\pi^{2} T}
\sum_f g_{{}_f} q_{{}_f}^{2} \int dp\frac{{p}^4}{E^2}\nonumber\\
 &\times&\left[\tau_{{}_f} f_{{}_f}^{0}(1-f_{{}_f}^{0})+ 
\tau_{\bar {{}_ f}}{\bar f_{{}_f}^{0}}
 (1-{\bar f_{{}_f}^{0}})\right]\nonumber\\
&-&\xi \frac{1}{12\pi^2 T^2} \sum_f g_{{}_f} q_{{}_f}^{2} \int dp\frac{{p}^6}{E^3}\\
&\times&\left[e^{(E-\mu^q_f)/T}\tau_{{}_f} (f_{{}_f}^{0})^2+ e^{(E+\mu^q_f)/T}
\tau_{\bar{{}_f}}({\bar f_{{}_f}^{0}})^2\right],\nonumber
\label{sigianiso}
\end{eqnarray}
where $f_{{}_f}^0$ and ${\bar f_{{}_f}^0}$ acquire the form as given in Eq. (16) and (17), respectively.
\subsection{Quasiparticle Model: Effective Masses and Relaxation Times}
Quasiparticles are the thermal excitations of the interacting
quarks and gluons retaining the quantum numbers of the real particles,
i.e., the quarks and gluons. In quasiparticle models~\cite{pe.1}, QGP
is described
by the system of ’massive’ noninteracting quasiparticles where the mass of
these quasiparticles
is temperature-dependent and arises because of the interactions of quarks
and gluons with the surrounding matter in the medium.
The effective mass of these quasiparticles is given by~\cite{pks}:
\begin{equation}
m_{{}_f}^{2}=m_{0,f}^{2}+\sqrt{2}m_{0,f}m_{th,f}+m_{th,f}^{2},
\end{equation}
where $m_{0,f}$ is the current mass of the flavour, $f$
and $m_{th,f}$ is the thermal mass of the flavour, $f$,
which is given by
\begin{equation}
m_{th,f}^{2}=\frac{g^{2}(T)T^{2}}{6}\left(1+\frac{(\mu^q_{{f}})^{2}}{\pi^{2}T^{2}}\right).
\end{equation}
Here $g^{2}$ is the QCD running coupling constant which in two-loop has following form~\cite{laine,agotiya} :
\begin{eqnarray}
\alpha_{S}(T,~\mu_q)&=&\frac{g^{2}(T,~\mu_q)}{4 \pi}=\frac{6 \pi}{\left(33-2 N_{f}
\right)\ln \left(\frac{T}{\Lambda_{T}}\sqrt{1+a\frac{\mu_q^{2}}{T^2}}\right)}\\
&\times&\left(1-\frac{3\left(153-19 N_f \right)}
{\left(33-2 N_f\right)^2}\frac{\ln \left(2 \ln \frac{T}{\Lambda_T}
\sqrt{1+a\frac{\mu_q^{2}}{T^2}} \right)}{\ln \left(\frac{T}{\Lambda_{T}}\sqrt{1+a\frac{\mu_q^{2}}{T^2}}\right) }\right),\nonumber
\end{eqnarray}
\noindent
where  $\Lambda_{T}$ is the QCD scale-fixing parameter which 
characterizes the strength of the interaction. It originates 
from the lowest non-zero Matsubara modes~\cite{vuo}. Here 
parameter $a$ is equal to $\frac{1}{\pi^{2}}$.\\

In Eqs. (\ref{sigiiso}) and (\ref{sigianiso}), $\tau_{f}$ is 
the collision time for which we use the following 
expressions for quarks (anti-quarks) from Ref.~\cite{hosoya}:
\begin{equation}
\tau_{q(\bar{q})}=\frac{1}{15\alpha_{s}^{2} T{ } log \left(\frac{1}{\alpha_{s}}\right)\left(1+0.06 N_{F}^{eff}\right)},
\end{equation}
\begin{figure}[!ht]
   \begin{center}
       \includegraphics[height=25em]{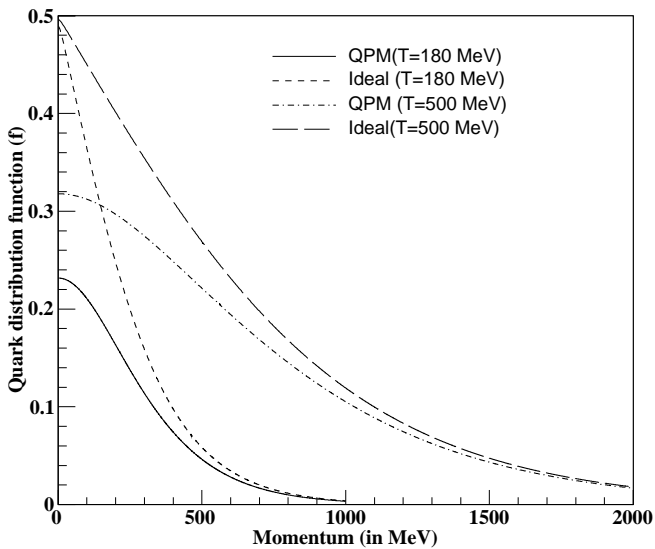}
  \label{Fig 1}
\end{center}
 \caption{Variation of QPM and ideal distribution function 
 with respect to momentum at two different temperatures. Solid and dash-dotted curve 
represents the light quark distribution function in QPM at $T=180$ and $500$ MeV, respectively.
Similarly short-dashed and long-dashed curve shows the corresponding results in ideal description.}
\end{figure}
\begin{figure}[!ht]
   \begin{center}
       \includegraphics[height=25em]{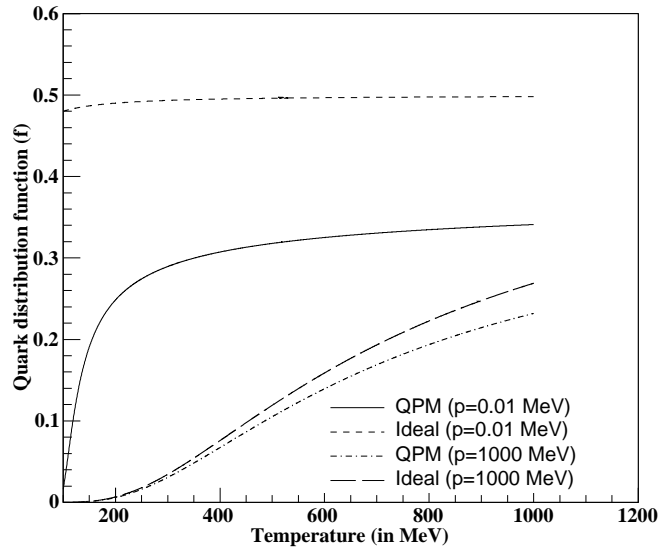}
  \label{Fig 2}
\end{center}
 \caption{Variation of QPM and ideal distribution functions 
 with respect to temperature at two different momentum.Solid and dash-dotted curve 
represents the light quark distribution function in QPM at $p=0.01$ and $1000$ MeV, respectively.
Similarly short-dashed and long-dashed curve shows the corresponding results in ideal description.}
\end{figure}
where $N_{F}^{\rm{eff}}$ is the number of effective flavour 
degrees of freedom. \\
Before going to the electrical conductivity results, we 
want to just provide the difference between the ideal 
description (where the mass consists of only the current mass) 
and the quasiparticle description (where the mass consists 
of current mass along-with a thermally generated mass) of QGP. 
To understand the crucial difference between these two 
descriptions, it is better to plot the occupation probability 
or distribution function in both descriptions. In Fig. 1, 
we have shown the variation of QPM distribution function for 
light quarks ($u$ and/or $d)$ with respect to momentum at two 
different temperatures $T=180$ and $500$ MeV. We further compare 
these results with the corresponding ideal distribution functions where we 
use only the current mass of the quarks in the mass term and 
there is no any thermal contribution. We observe that at low momentum the 
difference between QPM and ideal description is large. The 
QPM occupation probability is low in comparison to ideal 
case and thus we can indirectly say that the number density 
will also be small. However, at higher momentum, both
picture give the same distribution function. 
Similarly, Fig. 2 demonstrates the variation of QPM distribution 
function with respect to temperature at two different momentum 
$p=0.01$ MeV (low momentum) and $1000$ MeV (high momentum). 
We have also plotted the corresponding ideal distribution 
function for comparison. Here we observed that the QPM 
distribution function or QPM occupation probability is small 
in comparison to ideal case at low momentum and this 
difference increases as we move from higher temperatures 
towards lower temperature. However, at higher momentum the 
difference between QPM and ideal distribution function 
is quite insignificant over the entire temperatures range 
considered by us. As we know from non-relativistic Drude's formula that 
the electrical conductivity is directly proportional to the 
number density which is nothing but the integration of 
distribution function over momentum space at any particular 
temperature. Therefore one can understand the usefulness 
of QPM description in describing the thermodynamical and 
transport properties of QGP specially near the critical 
temperature ($T_{c}$). 
\section{Results and Discussions}
\begin{figure}[!ht]
   \begin{center}
       \includegraphics[height=25em]{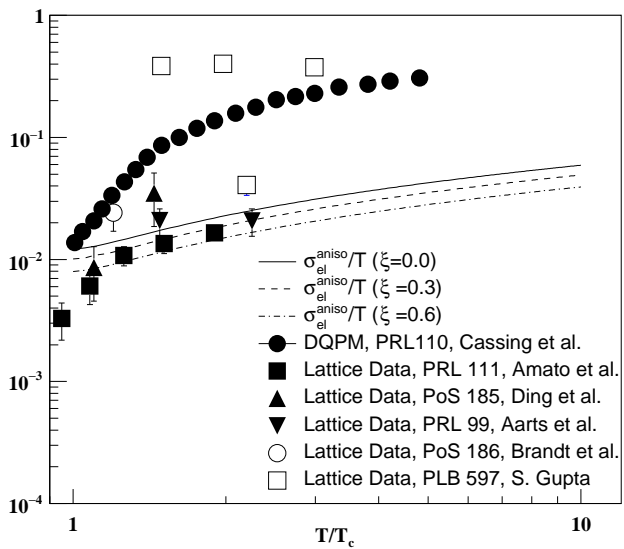}
  \label{Fig 3}
\end{center}
 \caption{Variation of ratio of electrical 
 conductivity to temperature ($\sigma_{\rm{el}}^{\rm{aniso}}/T$) with respect 
 to temperature at different values of anisotropy parameter 
 $\xi=0.0,~0.3$ and $0.6$ and they are shown in the plot by 
 solid, dashed and dash-dotted line, respectively. Green 
 points are the results as obtained from dynamical 
 quasiparticle model (DQPM) in Ref.~\cite{cassing}. 
 All other symbols are the data points obtained from 
 various lattice calculations~\cite{aamato,ding, aarts,gupta}.}
\end{figure}
In Fig. 3, we present the variation of ratio of electrical 
conductivity to temperature ($\sigma_{\rm{el}}^{\rm{aniso}}/T$) with respect 
to temperature at $\mu_q=0$ for an anisotropic QGP. 
For weakly anisotropic system, we choose three constant 
values of $\xi$ as $0.0,~0.3$ and $0.6$, where $\xi=0.0$ represents 
$\sigma_{\rm{el}}/T$ in isotropic case as mentioned in Eq. (11). 
We have compared our model results with the corresponding 
results obtained in various lattice calculations~
\cite{aamato,ding,aarts,gupta}. We further compared our 
model results with the $\sigma_{\rm{el}}/T$ obtained in DQPM 
model~\cite{cassing}. We observe that $\sigma_{\rm{el}}^{\rm{aniso}}/T$ 
increases monotonically with temperature starting from a 
lower value at $T=T_{c}$, where $T_{c}$ is the crossover 
temperature for the transition from QGP to hadron gas (HG). 
Further we observe that as the anisotropy increases from $0.0$ 
to $0.6$, the value of $\sigma_{\rm{el}}/T$ decreases for all the 
temperatures. $\sigma_{\rm{el}}^{\rm{aniso}}/T$ satisfies the lattice results 
well when the anisotropy parameter has a value equal to $0.6$. 
The results obtained from DQPM overestimate the value of 
$\sigma_{\rm{el}}/T$ as compared to the lattice as well as our model 
calculations. However, we cannot say at this moment the exact 
status of any model calculations since the lattice results 
are distributed over a wide range. 
\begin{figure}[!ht]
    \begin{center}
       \includegraphics[height=25em]{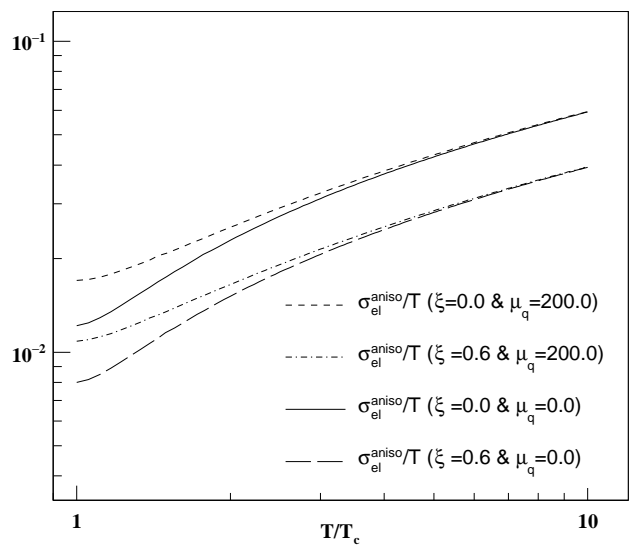}
  \label{fig:Fig 4}
\end{center}
 \caption{Variation of ratio of electrical 
 conductivity to temperature ($\sigma_{\rm{el}}^{\rm{aniso}}/T$) with respect 
 to temperature at two different values of anisotropy parameter 
 $\xi=0.0$ and $0.6$ for quark chemical potential $\mu_q=0$ and $\mu_q=200~MeV$.}
\end{figure}
\\

Fig.4 represents the variation of $\sigma_{\rm{el}}^{\rm{aniso}}/T$ with 
respect 
to temperature at quark chemical potential $\mu_q=200~MeV$ and compare them 
with our model results at $\mu_q=0 ~MeV$. Here we have 
shown the results for two different values of $\xi$ equal to 
$0.0$ and $0.6$. Solid and long-dashed curve demonstrate the $\sigma_{\rm{el}}^{
\rm{aniso}}$
at $\mu_q=0.0$ for $\xi=0.0$ and $0.6$, respectively. Short-dashed and dash-dotted curve
show the results at $\mu_q=200$ MeV for $\xi=0.0$ and $0.6$, respectively. At large temperatures we found that 
$\sigma_{\rm{el}}^{\rm{aniso}}/T$ at finite $\mu_q$ remains same as in the case 
of zero $\mu_q$. However, the 
value of $\sigma_{\rm{el}}^{\rm{aniso}}/T$ becomes large for finite $\mu_q$ in comparison 
to the value at $\mu_q=0 ~MeV$ as the temperature decreases below $T=4T_{c}$. 
This behaviour is quite consistent with the results obtained in a 
very recent DQPM calculations at finite $\mu_q$~\cite{berre}.\\

In summary we have studied the behaviour of electrical 
conductivity for QGP phase in the framework of relativistic 
Boltzmann's kinetic equation using relaxation time approximation. 
In this work we describe the QGP as a system of quasiparticles 
having temperature and chemical potential dependent mass 
along with their rest masses. Firstly we derive the expression 
to calculate $\sigma_{\rm{el}}$ in an isotropic medium. Later a momentum 
anisotropy in the distribution function of quarks and anti-quarks
has been introduced and taking the leading order contribution 
in distribution function we have obtained $\sigma_{\rm{el}}$  in 
an anisotropic medium which is equal to the $\sigma_{\rm{el}}^{\rm{iso}}$ 
minus the correction factor due to momentum anisotropy. 
We have plotted $\sigma_{\rm{el}}^{\rm{aniso}}/T$ for different values of 
$\xi$ which is equal to $0.0,~0.3$ and $0.6$. We have compared 
our model results with the corresponding results obtained in 
various lattice results and found a reasonable agreement between 
them for certain value of $\xi$. Further, we have plotted our model 
expectations for $\sigma_{\rm{el}}^{\rm{aniso}}/T$  at $\mu_q=200~MeV$ and compare 
them with the model results obtained at zero chemical potential. 
Both in isotropic as well as in anisotropic case, we find the similar 
behaviour as observed in DQPM calculations~\cite{berre} i.e., increase in the
value of $\sigma_{\rm{el}}/T$ with increase in quark chemical potential near $T_{c}$.
 At very high temperatures ($T\sim 8-10 T_{c}$), the difference 
between $\mu_q=0.0$ and $\mu_q=200~MeV$ case is very small. These 
calculations are done in a static system when there is no any 
proper time dependence has been given to the anisotropy parameter. 
However, in realistic situation, $\xi$ varies with the proper time 
starting from the initial proper time up to a time when the system 
becomes isotropic and $\xi$ becomes zero. Thus one has to incorporate 
a proper time 
dependence to the anisotropy parameter. Work in this direction is 
in progress and will be presented elsewhere. 
 
\noindent
\section{Acknowledgments}
The authors are thankful for financial assistance from Council of Scientific and Industrial Research (No. CSR-656-PHY), Government of India.

\newpage

\begin{thebibliography}{99}
\bibitem{roma} P. Romatschke and U. Romatschke, Phys. Rev. Lett. {\bf 99}, 172301 (2007); B. Schenke, S. Jeon and C. Gale, Phys. Rev. {\bf C 82}, 014903 (2010).
\bibitem{heinz} U. Heinz, P.F. Kolb, Nuclear Phys. {\bf 702} (2002).
\bibitem{kotun}P. K. Kovtun, D. T. Son and A. O. Starinets, Phys. Rev. 
Lett. {\bf 94}, 111601 (2005).
\bibitem{lee} T.D. Lee, Nuclear Phys. {\bf 750}, 1 (2005).
\bibitem{guy} M. Gyulassy, L. Mclerran, Nuclear Phys. {\bf 750}, 30 (2005).
\bibitem{shuryak} E.V. Shuryak, Nuclear Phys. {\bf 750}, 64 (2005).
\bibitem{hirano} T. Hirano, M. Gyulassy, Nuclear Phys. {\bf 769}, 71 (2006).
\bibitem{vgreco} A. Puglisi, S. Plumari and V. Greco, arXiv:1407.2259v1[hep-ph] (2014).
\bibitem{fraile} D. Fernandez-Fraile and A. Gomez Nicola, Phys. Rev. D {\bf 73}, 045025 (2006).
\bibitem{greif} M. Greif, I. Bouras, C. Greiner, Z. Xu, Phys. Rev. D {\bf 90}, 094014 (2014).
\bibitem{tuchin} K. Tuchin, Advances in High Energy Physics {\bf 2013}, 490495 (2013).
\bibitem{fuku} K. Fukushima, D. E. Kharzeev and H. J. Warringa, Phys. Rev. D {\bf 78}, 074033 (2008).
\bibitem{hongo} Y. Hirono, M. Hongo, T. Hirano, arXiv:1211.1114[hep-ph] (2012).
\bibitem{kapusta} J. Kapusta, Finite-temperature field theory, Cambridge monographs on mathematical physics, Cambridge University Press (1993).
\bibitem{turbide} S. Turbide, R. Rapp, and C. Gale, Phys. Rev. C {\bf 69}, 014903 (2004).
\bibitem{linnyk} O. Linnyk, W. Cassing and E. Bratkovskaya, Phys. Rev. C {\bf 89}, 034908 (2014).
\bibitem{strick} M. Strickland, arXiv:1401.1188v1[nucl-th] (2014).
\bibitem{strick1} M. Strickland, Nucl. Phys. A {\bf 926}, 92 (2014); arXiv: 1312.2285[hep-ph] (2013).
\bibitem{heller} M. P. Heller, R. A. Janik, and P. Witaszczyk, Phys. Rev. Lett. {\bf 108}, 201602 (2012).
\bibitem{vander} W. vander Schee, P. Romatschke, and S. Pratt, Phys. Rev. Lett. {\bf 111}, 222302 (2013).
\bibitem{mac} L. D. McLerran, and R. Venugopalan, Phys. Rev. D {\bf 49}, 2233 (1994).
\bibitem{iancu} E. Iancu and R. Venugopalan, hep-ph/0303204 (2003).
\bibitem{pks} P. K. Srivastava, S. K. Tiwari and C. P. Singh, Phys. Rev. D {\bf 82}, 014023 (2010).
\bibitem{pks1} P. K. Srivastava and C. P. Singh, Phys. Rev. D {\bf 85}, 114016 (2012).
\bibitem{aamato} A. Amato, G. Aarts, C. Allton, P. Giudice, S. Hands, and J. Skullerud, Phys. Rev. Lett. {\bf 111}, 172001 (2013).
\bibitem{ding} H. -T. Ding, A. Francis, O. Kaczmarek, F. Karsch, E. Laermann, W. Soeldner, Phys. Rev. D {\bf 83}, 034504 (2011).
\bibitem{aarts} G. Aarts, C. Allton, J. Foley, S. Hands, S. Kim, Phys. Rev. Lett. {\bf 99}, 022002 (2007).
\bibitem{gupta} S. Gupta, Phys. Lett. B {\bf 597}, 57 (2004).
\bibitem{berre} H. Berrehrah, E. Bratkovskaya, W. Cassing and R. Marty, arXiv:1412.1017v1[hep-ph] (2014).
\bibitem{Yagi} K. Yagi, T. Hatsuda, Y. Miake, \emph{Quark-Gluon Plasma: from big bang to little bang}, Cambridge University Press, 2005.
\bibitem{Cercignani} C. Crecignani and G. M. Kremer, \emph{The Relativistic Boltzmann Equation: Theory and Applications}, Boston; Basel; Berlin: Birkhiiuser, 2002.
\bibitem{green} M. S. Green, J. Chem. Phys. {\bf 20}, 1281 (1952).
\bibitem{kubo} R. Kubo, J. Phys. Soc. Jpn. {\bf 12}, 570 (1957).
\bibitem{dum1} A. Dumitru, Y. Guo and M. Strickland, Phys. Lett. B {\bf 662}, 37 (2008).
\bibitem{dum2} A. Dumitru, Y. Guo, A. Mocsy and M. Strickland, Phys. Rev. D {\bf 79}, 054019 (2009).
\bibitem{gribov} L. V. Gribov, E. M. Levin and M. G. Ryskin, Phys. Rept. {\bf 100}, 1 (1983).
\bibitem{mueller} A. H. Mueller and J.-W. Qiu, Nucl. Phys. B {\bf 268}, 427 (1986).
\bibitem{blaiz} J. P. Blaizot and A. H. Mueller, Nucl. Phys. B {\bf 289}, 847 (1987).
\bibitem{roma1} P. Romatschke and M. Strickland, Phys. Rev. D {\bf 68}, 036004 (2003).
\bibitem{roma2} P. Romatschke and M. Strickland, Phys. Rev. D {\bf 70}, 116006 (2004). 
\bibitem{Martinez:2010sd-12tu} M. Martinez and M. Strickland,  Nucl. Phys. A {\bf 856}, 68 (2011);  ~Nucl. Phys. A~{\bf 848}, 183  (2010)
\bibitem{Martinez:PRC852012}M. Martinez, R. Ryblewski,  and M. Strickland, Phys. Rev. C {\bf 85}, 064913 (2012).
\bibitem{Ryblewski:2010bs} R. Ryblewski, and W. Florkowski, J. Phys. G {\bf 38}, 015104   (2011);~ Euro. Phys. J. C {\bf 71}, 1761 (2011).
\bibitem{Ryblewski:2012rr}R.~Ryblewski, W.~Florkowski, Phys. Rev. C {\bf 85}, 064901 (2012);W.~ Florkowski, R.~Ryblewski, and M.~Strickland,  Phys. Rev. D {\bf 86}, 085023 (2012).
\bibitem{Florkowski:2010cf} W.~Florkowski and R.~Ryblewski,  Phys. Rev. C
{\bf 83}, 034907 (2011).
\bibitem{lata:PRD2013} L. Thakur, N. Haque, U. Kakade, and Binoy Krishna Patra, 
Phys. Rev.  D {\bf 88}, 054022 (2013).
\bibitem{lata:PRD2014} L. Thakur, U. Kakade, and Binoy Krishna Patra, 
Phys. Rev.  D {\bf 89}, 094020 (2014).
\bibitem{pe.1} A. Peshier, B. Kampfer, O. P. Pavlenko and G. Soff, Phys. 
Lett. B {\bf 337}, 235 (1994)
\bibitem{asa} M. Asakawa, S. A. Bass, and B. Muller, Prog. Theor. Phys. {\bf 116}, 725 (2007).
\bibitem{laine} M. Laine, Y. Schroder, J. High. Energy Phys. {\bf 0503}, 067 (2005).
\bibitem{agotiya} V. Agotiya, L. Devi, U. Kakade and B. K. Patra, Int. J. Mod. Phys. A {\bf 1250009} (2012).
\bibitem{vuo} A. Vuorinen, arXiv:hep-ph/0402242.
\bibitem{hosoya} A. Hosoya and K. Kajantie, Nucl. Phys. B {\bf 250}, 666 (1985).
\bibitem{cassing} W. Cassing, O. Linnyk and T. Steinert, Phys. Rev. Lett. {\bf 110}, 182301 (2013).
 \end{thebibliography}
\end{document}